\documentclass[conf]{new-aiaa}
\usepackage[utf8]{inputenc}

\usepackage{graphicx}
\usepackage{amsmath}
\usepackage[version=4]{mhchem}
\usepackage{siunitx}
\usepackage{longtable,tabularx}
\title{Development of Techniques Enabling Suborbital Small Object Capture Around Low Gravity Asteroids}

\author{Leonard D. Vance\footnote{PhD Candidate, Aerospace and Mechanical Engineering, University of Arizona, 1130 N Mountain Ave, Tucson, AZ 85721} and Jekan Thangavelautham\footnote{Assistant Professor, Aerospace and Mechanical Engineering, University of Arizona, 1130 N Mountain Ave, Tucson, AZ 85721}}

\begin{document}

\maketitle

\begin{abstract}
The exploration of small body asteroids provides direct access to the primitive building blocks of our solar system.  Sample and return missions enable chemical and radioisotope studies which not only provide evidence for the formation of the solar system, but also a basic understanding of where resources might be found for future exploitation.  The touch-down and sample techniques established by Hayabusa-2 and OSIRIS-REx accomplish this mission by physically touching down on the asteroid and collecting samples into a basket extended via a probe from the bottom of the spacecraft.  This technique has been demonstrated to work, but contains a high cost in both mission operations as well as the size and complexity of the collection mechanism itself.  This paper explores an alternative sample and return technique by exploiting the recent discovery of regolith particle ejections from Bennu.  Particles ejected from the surface of Bennu are typically 1 cm in size and spend several hours in flight, suggesting the possibility that nanospacecraft deployed from the mother-spacecraft could chase down, collect and return with the sample with minimal sensor and delta-V capability.  Key aspects of this mission are developed to reduce risk, and an overall mission concept is developed to establish plausibility.
\end{abstract}

\section{Introduction}
Asteroid sample and return missions provide scientists with direct access to materials formed during contraction of our original solar planetary disc.  These missions will also play an important part in resource prospecting, especially for water which can be used a propellant for interplanetary spacecraft~\cite{65}.   Chemical and isotopic analysis of these samples provide the gold standard for understanding the age, history and evolution of the overall solar system.  Several missions have therefore attempted this, including Hayabusa 1 and 2, OSIRIS-REx, and NASA’s Stardust, but the challenges involved in gathering surface samples has proven more difficult than perhaps originally expected.  The combination of very low gravity, relatively high spin rates, and the unpredictability of the surface composition have proven difficult to predict, and the results from initial missions have been mixed.  Improvements in retrieval techniques could plausibly reduce mission risks and costs significantly. Mother-Daughter architectures have also been explored, exploring the plausibility of daughter vehicles to undertake local missions in swarms which would would not be plausible with a single spacecraft~\cite{64}. This also avoid a single daughter craft having to have long-range communications such as using inflatable antennas~\cite{66}.

\begin{figure}
\includegraphics[width=\linewidth]{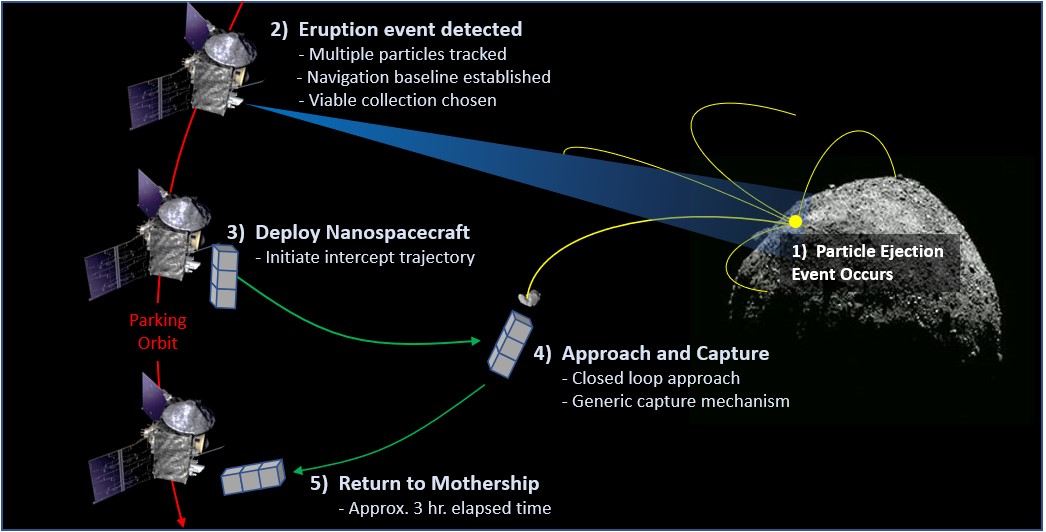}
\caption{Aspects of the proposed mission.  Nano-spacecraft are deployed from a parking orbit to collect ejected particles and return to the mother spacecraft}
\label{intro_fig1}
\end{figure}

In December 2018, OSIRIS-Rex arrived at the near-earth asteroid Bennu as a prelude to touching down and retrieving such a sample.  The initial survey of the asteroid provided two significant surprises.  The  first involved the lack of fine particulate matter on the surface, and the second was the discovery that Bennu is a “live” asteroid, with low velocity objects erupting regularly from the surface.  The first discovery provides a significant challenge to the mission as the sample system is designed to collect relatively small particles, while the second suggests that future sample and returns could use small deployed spacecraft, capturing the expelled samples mid-flight before returning to the mother spacecraft.  

This paper explores the design of such a mission, which is plausibly less complex than the sample and return techniques currently utilized.  Techniques developed for such a mission would also enable other high value missions such as low earth orbit debris capture and potential on-orbit servicing by nano-spacecraft.

Several areas of novel research must be explored to develop a plausible system design.  To begin with, characteristics of the expelled objects must be established.  Positions, velocities, sizes and eruption frequencies are needed. It is credible that the movement of these objects in part explain the “double-top” shape of spinning rubble pile asteroids, so it may be conversely possible that the “double-top” shape implies the existence of object expulsions, and therefore indicates opportunities to collect samples without touching the surface.

The design of a temporarily deployed nano-spacecraft to capture expelled objects provides significant technical challenges.  Each vehicle must be able to navigate, guide, collect, return and dock with the parent spacecraft within the timeframe of a typical particle trajectory.  To begin with, autonomous navigation can be established using passive line of sight histories to the targeted object.  The inertial line of sight history between two orbital objects provides sufficient information to establish estimate ephemeris for both objects, and this can be realized with a modified Extended Kalman Filter (EKF) formulation, possibly simplified to a constant gain model to reduce processor utilization.  The resulting implementation of closed loop guidance using the aforementioned navigation output with Lambert’s method is well known but computationally expensive, and is thus replaced with research towards a perceptron based network trained for operation within any particular asteroid’s gravitational field.  

Spacecraft control involves both attitude and position control, and would be exercised via a six degree of freedom thruster system possibly including reaction wheels for fine control.  The concept of a minimal redundant thruster configuration for nanospacecraft is explored via null vector analysis, leading to candidate minimal thruster systems with single failure redundancy.  Finally, a generic grappling system is discussed for a non-cooperative target, with possible application to wider missions such as debris removal from low earth orbit.  The overall proposed mission is summarized in Figure~\ref{intro_fig1}.

\section{Ejected particle characterization on Bennu}
The process of rubble pile asteroid formation was discussed by Michel~~\cite{13} in 2001, and Guibot and Scheeres~~\cite{11} in 2003, but innovations in deep space radar imaging led to the confirmation in 2006~~\cite{17},~\cite{23} that asteroid (66391) 1999 KW4 has a top-shape with a nearly constant slope of about 35 degrees from mid latitudes down to the equator. It is a fast-rotator, with a period 2.8 hr and a small moon just off the equator suggesting recent mass shedding~~\cite{23}. Subsequent radar observations of (341843) 2008 EV5~~\cite{18}, (29075) 1950 DA~~\cite{16} and (10195) Bennu~~\cite{24} show that top-shapes are common for near-earth asteroids. Arrival of OSIRIS-REx and Hayabusa-2 confirmed this~~\cite{31} for both (10195) Bennu and (162173) Ryugu~~\cite{9} respectively, providing detailed 3D models to enable detailed analysis

Centripetally-induced landslides caused by solar-pressure induced YORP spin-up have been proposed to explain current observations~~\cite{11},~~\cite{8}. As more data have become available, these ideas have been developed analytically~~\cite{2},~\cite{3},~\cite{6},~\cite{10},~\cite{22} and avalanche failure mechanisms have been explored~~\cite{29}, ~\cite{27}. Particle models~~\cite{19},~\cite{26},~\cite{30}, some including cohesion~~\cite{14},~\cite{15}, have been applied to the problem.  There have also been attempts to simulate low speed collisions as a possible origin for top-shapes. Simulations of impacts between two near equal mass rubble piles, where friction and porosity is included~~\cite{5}, can produce final populations where top-shapes are common. Additionally, the direct measurements of the YORP induced angular acceleration on Bennu~~\cite{1},~\cite{20}, and its sensitivity to variations in configuration~~\cite{21} provide an understanding of the possible chaotic nature of Bennu's accelerating spin rate.

\begin{figure}
\includegraphics[width=\linewidth]{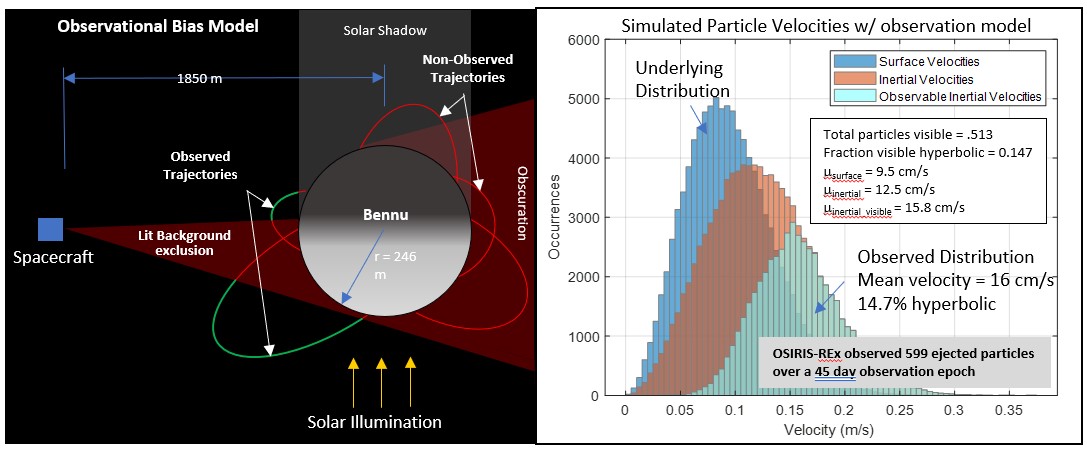}
\caption{Particle frequency and velocity distributions derived from OSIRIS-REx observations provide a basis for the proposed mission}
\label{fig1}
\end{figure}

The discovery in 2019 by NASA's OSIRIS-REx mission that the 500-metre diameter carbonaceous rubble pile Bennu is an active asteroid, regularly ejecting objects from the surface~~\cite{4}, suggests the possibility that objects can be captured in flight without approaching the asteroid surface.  Some 599 objects were observed by OSIRIS-REx over a 45 day observation period following its initial approach to the asteroid, providing a mean of 13 objects observed per day.   When a simulation experiment is executed accounting for the obscuration of most of the actual particles leaving the surface, the actual launch frequency is about 26 particles per day with a mean launch velocity of 9.5 cm/s as shown in Figure~\ref{fig1}.  The resulting flight time for an average particle is about 1.5 hours given Bennu's very low gravity.  This ensures that particles will be available for sampling on a near daily basis.

\section{Own-Ship and Target Navigation with a Single Passive Line-Of-Sight Sensor}
Navigation techniques using passive line-of-sight measurements are well known within the guidance, navigation and control community.  The ability to infer ephemeris information of an orbital object by tracking it from a known position is well understood, as is the ability to infer one's own ephemeris by tracking an object with known position and velocity.  Here we show that it is possible to infer both own-ship and tracked-object ephemeris given knowledge of the gravitational influences both are subjected to.

\begin{figure}
\includegraphics[width=\linewidth]{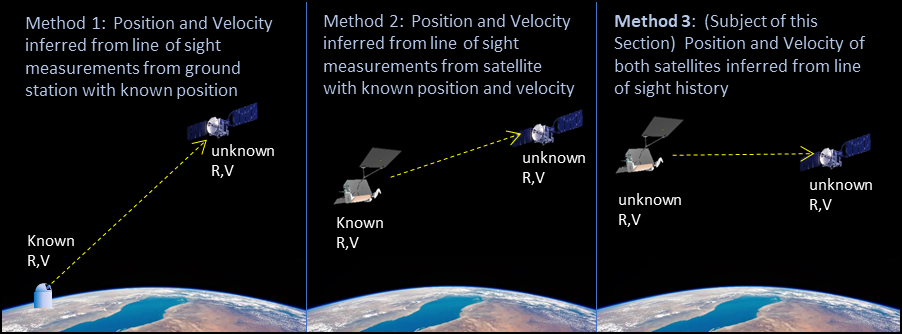}
\caption{The line of sight history between two objects under the influence of a common gravitational body is sufficient to estimate the position and velocity of both}
\label{autonav_fig1}
\end{figure}

Figure~\ref{autonav_fig1} shows the general nature of the problem.  It is well established that position and velocity data for an overhead satellite can be extracted from passive line of sight measurements from the ground.  Observability of range is inferred from the gravitational attraction of the orbiting object.  Likewise, as a simple extension, it is also possible to do the same job from an orbiting satellite.  If you know your own position and velocity, you can infer the position and velocity of another satellite given a time history line of sight measurements. 

With some additional thought, it is possible that the position and velocity of both the tar-get and observer satellites can be inferred from a single line of sight history.  With some exceptions, it can be asserted that any line of sight history between two objects gravitation-ally influenced by a common source is the result of only one specific target and home satellite trajectory.

\subsection{Method}
The Extended Kalman Filter (EKF) is a well-known, pseudo-optimal predictor-corrector filter widely used for a wide variety of aerospace purposes.  This section shows the adaptation of the basic filter structure for this particular problem.  The established form of the EKF~~\cite{32}  starts by propagating the current estimate and covariance matrices forward by one timestep.

\begin{equation}
    \boldsymbol{\hat{x}}(k+1|k) = \boldsymbol{\hat{x}}(k|k) + \int_{t_k}^{t_{k+1}}\boldsymbol{f}[\hat{x}(t|t_k),\boldsymbol{u}^*(t),t]dt
\end{equation}

\begin{equation}
    P(k+1|k) = \Phi(k+1|k)P(k|k)\Phi^{'}(k+1,k)+Q_d(k)
\end{equation}
 
This followed by calculation of the Kalman gains:

\begin{equation}
    \boldsymbol{K}(k+1)=\boldsymbol{P}(k+1|k)[\boldsymbol{H}^{'}_x(k+1)*\boldsymbol{P}(k+1|k)\boldsymbol{H}^{'}_x(k+1)+\boldsymbol{R}(k+1)]^{-1}
\end{equation}

 The state vector estimate can then be updated:
 
 \begin{equation}
 \hat{x}(k+1|k+1)=\hat{x}(k+1|k)+\boldsymbol{K}(k+1)\{\boldsymbol{z}(k+1)-\boldsymbol{h}[\boldsymbol{\hat{x}}(k+1|k),\boldsymbol{u}^{*}(k+1),k+1]-\boldsymbol{H}_{u}(k+1)\delta\boldsymbol{u}(k+1)\}
 \end{equation}
 
 Finally, also update the covariance matrix:
 \begin{equation}
     \boldsymbol{P}(k+1|k+1) = [\boldsymbol{I}-\boldsymbol{K}(k+1)\boldsymbol{H}_x(k+1)]\boldsymbol{P}(k+1|k)
 \end{equation}
 
The definition of these parameters are consistent with normal EKF usage:\\

     \hspace*{1cm} $\boldsymbol{\hat{x}}$ = state estimate\\
     \hspace*{1cm} $k$ = current time step\\
     \hspace*{1cm} $k+1$ = next time step\\
     \hspace*{1cm} $\boldsymbol{f}$ = function propagating state estimate $\hat{\boldsymbol{x}}$ in time\\
     \hspace*{1cm} $t$ = time\\
     \hspace*{1cm} $\boldsymbol{u}^*$ = nominal input\\
     \hspace*{1cm} $\boldsymbol{P}$ = Covariance matrix\\
     \hspace*{1cm} $\boldsymbol{\Phi}$ = Linearized time propagation matrix\\
     \hspace*{1cm} $\boldsymbol{Q_d}$ = State error propagation matrix\\
     \hspace*{1cm} $\boldsymbol{K}$ = Kalman gain matrix\\
     \hspace*{1cm} $\boldsymbol{h}$ = function taking state vector to measurements\\
     \hspace*{1cm} $\boldsymbol{H_x}$ = Jacobean of $\boldsymbol{h}$ w/r to state vector $\boldsymbol{\hat{x}}$\\
     \hspace*{1cm} $\boldsymbol{H_u}$ = Jacobean of $\boldsymbol{h}$ w/r to external inputs\\
     \hspace*{1cm} $\boldsymbol{I}$ = Identity matrix\\

The function $\boldsymbol{u^*}$ and its derivatives are zero for the purposes of this exercise since there are no external forces acting on the system, and as such, (1) through (5) can be simplified to give

\begin{equation}
    \boldsymbol{\hat{x}}(k+1|k) = \boldsymbol{\hat{x}}(k|k) + \int_{t_k}^{t_{k+1}}\boldsymbol{f}[\hat{x}(t|t_k),t]dt
\end{equation}

\begin{equation}
    P(k+1|k) = \Phi(k+1|k)P(k|k)\Phi^{'}(k+1,k)+Q_d(k)
\end{equation}
 
for the propagation equations, and
 
 \begin{equation}
    \boldsymbol{K}(k+1)=\boldsymbol{P}(k+1|k)[\boldsymbol{H}^{'}_x(k+1)*\boldsymbol{P}(k+1|k)\boldsymbol{H}^{'}_x(k+1)+\boldsymbol{R}(k+1)]^{-1}
\end{equation}

\begin{equation}
    \hat{x}(k+1|k+1)=\hat{x}(k+1|k)+\boldsymbol{K}(k+1)\{\boldsymbol{z}(k+1)-\boldsymbol{h}[\boldsymbol{\hat{x}}(k+1|k)\}
 \end{equation}
 
 \begin{equation}
     \boldsymbol{P}(k+1|k+1) = [\boldsymbol{I}-\boldsymbol{K}(k+1)\boldsymbol{H}_x(k+1)]\boldsymbol{P}(k+1|k)
 \end{equation}
 
for the update equations.

The state estimation vector  contains 3 element position, velocity and accelerations for each object.  Since we are tracking one object from a home vehicle, there are 18 elements in the state estimation vector comprising:

\begin{figure}
    \includegraphics[width=\linewidth]{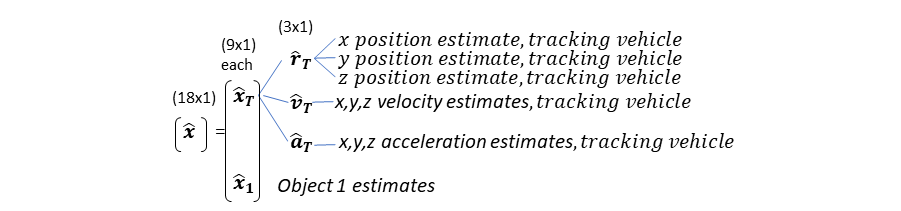}
    \caption{The line of sight history between two objects under the influence of a common gravitational body is sufficient to estimate the position and velocity of both}
    \label{autonav_fig2}
\end{figure}

The function $f$ is a simple orbital propagator, utilizing the basic gravitational law 

\begin{equation}
    \boldsymbol{a} = -\frac{G\,m_{bennu}\boldsymbol{r}}{|\boldsymbol{r}|^3}
\end{equation}

 where the r’s are taken from the position estimates in the state vector.  The corresponding matrix $\Phi$  is the Jacobean of this with respect to $\boldsymbol{\hat{x}}$ , giving an 18 × 18 matrix.  The function $\boldsymbol{h}$ takes state variables and constructs the measurements $\boldsymbol{z}$, and is therefore of the form:
 
\begin{equation}
    \boldsymbol{z} = \boldsymbol{h}(\boldsymbol{\hat{x}})
\end{equation}

Since the measurements for this system are the normalized line of sight vectors from the tracking vehicle to the two asteroids, the resulting function $\boldsymbol{h}$ is 

\begin{equation}
    \boldsymbol{h} = 
    \begin{bmatrix}
      \frac{x-x_t}{\sqrt{(x-x_t)^2+(y-y_t)^2+(z-z_t)^2}} \\
      \frac{x-x_t}{\sqrt{(x-x_t)^2+(y-y_t)^2+(z-z_t)^2}} \\
      \frac{x-x_t}{\sqrt{(x-x_t)^2+(y-y_t)^2+(z-z_t)^2}} \\
    \end{bmatrix}
\end{equation} 

The Jacobian of this function with respect to each component of the state estimate vector   provides the resulting 6 × 18 $\boldsymbol{H_x}$ matrix.

The $\boldsymbol{R}$ matrix represents the system measurement noise, and this takes the form of angular uncertainties along the line of sight to the object.  The measurement vector $\boldsymbol{z}$ is the normalized line of sight vector the target as given by an imaging system, so given an angular uncertainty $\epsilon$ in each axis, the covariance matrix for measurement errors to the tracked object, in line of sight coordinates is:

\begin{equation}
    \boldsymbol{R_{los}} = 
    \begin{bmatrix}
    1 & 0 & 0\\
    0 & \epsilon^2 & 0\\
    0 & 0 & \epsilon^2\\
   \end{bmatrix}
\end{equation}
 
This must then be rotated into an inertial coordinate frame used by the rest of the filter.  Derivation of the coordinate transform matrix is done by establishing sequential coordinate frame rotation angles about the z and then y-axes respectively, providing a coordinate transformation matrix of 

\begin{equation}
    \boldsymbol{T_{los}} = 
    \begin{bmatrix}
    cos\psi cos\theta  & sin\psi cos\theta  &  -sin\theta\\
    -sin\psi           & cos\psi            &  0         \\
    cos\psi sin\theta  & sin\psi sin\theta  &  cos\theta \\
   \end{bmatrix}
\end{equation}

Where the sequential angles of rotation about z and y axes respectively are:

\begin{equation}
    \psi = tan^{-1}(\frac{y-y_T}{x-x_T})
\end{equation}

\begin{equation}
    \theta = tan^{-1}(\frac{z-z_t}{\sqrt{(x-x_t)^2 + (y-y_T)^2}})
\end{equation}
 
This matrix takes a vector from inertial to the line of sight coordinated frame with respect to object 1.  Using a similarity transformation to convert from line of sight to inertial coordinates, we have 

\begin{equation}
    \boldsymbol{R}_{inertial}= \boldsymbol{T}_{los}\boldsymbol{R}_{los}\boldsymbol{T}^{T}_{los}
\end{equation}
 
Finally, the 18x18 state process noise matrix Q is constructed with arbitrarily small diagonal elements for the purpose of establishing feasibility.

\begin{equation}
    \boldsymbol{Q}= 1.0x10^{-30}*\boldsymbol{I}_{18x18}
\end{equation}

\begin{figure}
\includegraphics[width=\linewidth]{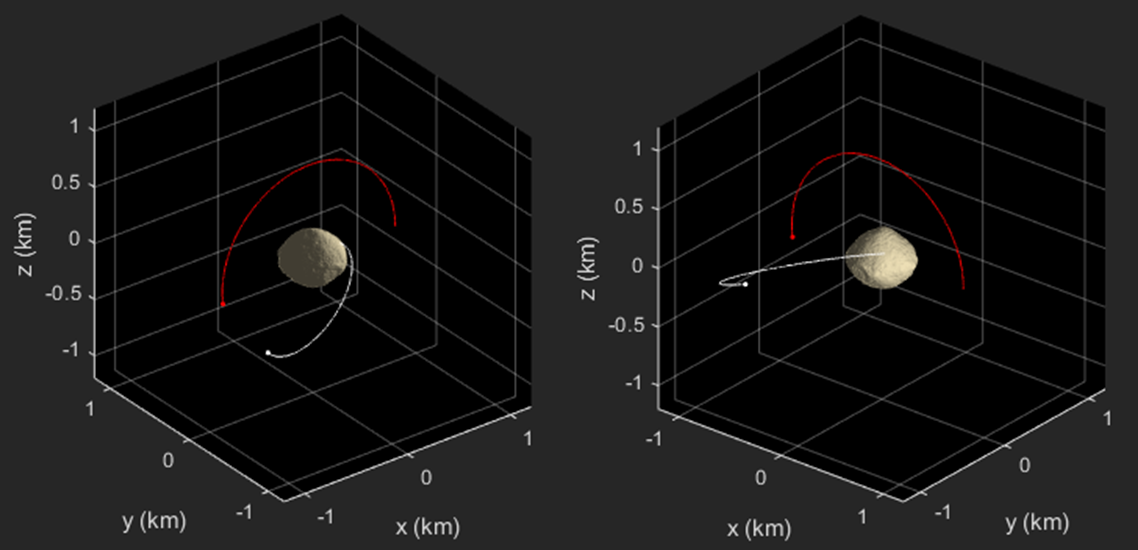}
\caption{Proof of concept trajectories for both the spacecraft and the ejected particle around (101955) Bennu}
\label{autonav_fig3}
\end{figure}

\subsection{Results}
The implementation of the above filter is tested in an environment derived from the physical environment of (101955) Bennu.  The specific intent is to show whether autonavigation can be performed from a nanospacecraft that is tasked with capture of an ejected particle from that surface.  As proof of concept, the spacecraft is placed in a circular orbit with an 800m semi-major axis with 90 degrees inclination.  The tracked object ejects from the mid latitudes of Bennu with a velocity of 17.5 cm/s, and it tracked by the spacecraft for a 10 hour epoch.  The trajectories of both objects relative to Bennu is shown in Figure~\ref{autonav_fig3}

\begin{figure}
\includegraphics[width=\linewidth]{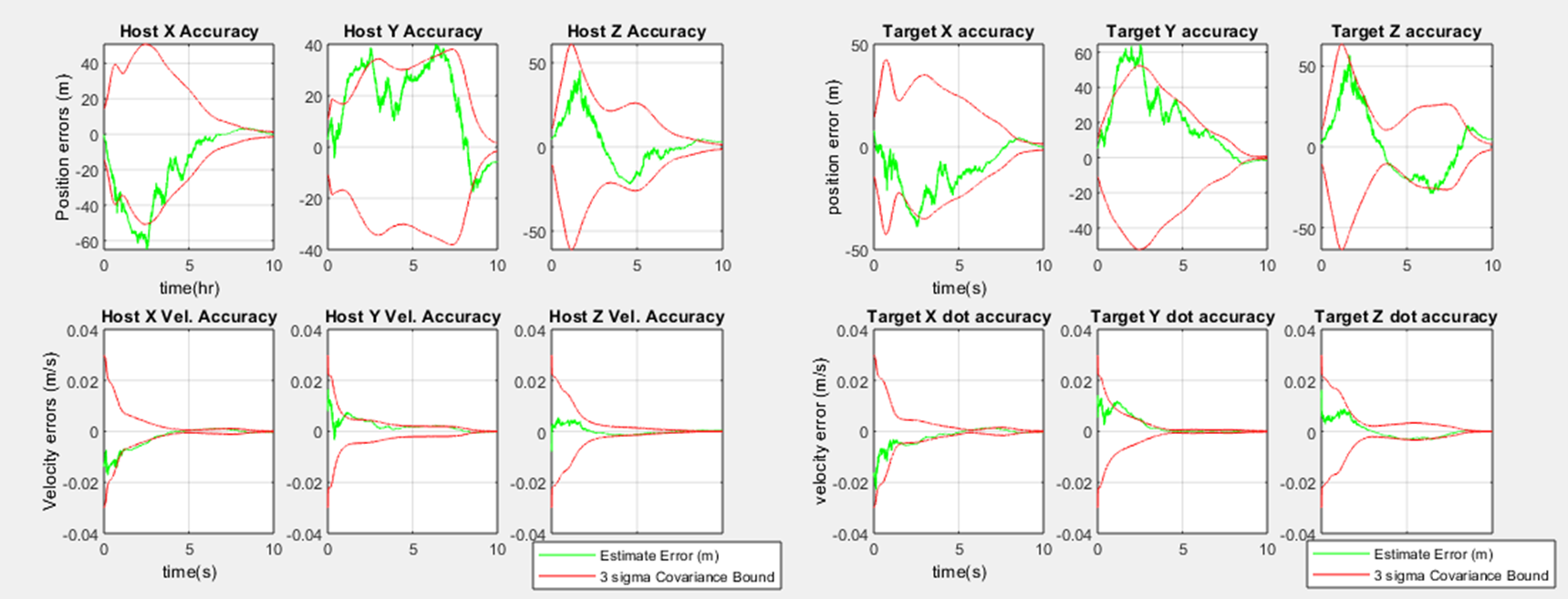}
\caption{Convergence of position and velocity estimates for both spacecraft and ejected particle over the 10 hour epoch described by the previous figure}
\label{autonav_fig4}
\end{figure}

Using only inertial line of sight measurements from the spacecraft to the ejected particle, position and velocity can be established for both simultaneously.  Given a reasonable initialization for covariances, the results are shown in Figure~\ref{autonav_fig4}, illustrating good convergence in position and velocity in all 3 axes, although over a time period of several hours. 

%
%
%
\section{Perceptron Based Orbital Guidance in a Low Gravity Asteroid Environment}
\subsection{Introduction}
Lambert's solution is a well studied approach~~\cite{46},~\cite{47} to orbital guidance, providing a near universal iterative solution for transition between two points under the influence of a single gravitational body.  The user specifies starting and end points along with transit time and central body gravitational parameter $\mu_m$, and the algorithm produces the corresponding velocities at those starting and end locations along with minimum and maximum ranges to the gravitational body.  While the algorithm is well known, accurate and almost universal in convergence, it requires a significant number of trigonometric function calls in an iterative framework.  In comparison, a feed forward neural network has low computational cost implementation and can theoretically mimic the output of much more complex algorithms if trained correctly.

Neural Network guidance could therefore be used by nanospacecraft to  rendezvous and gather the particles emitted by an active asteroid in flight. The discovery in 2019 that Bennu is an active asteroid, and is regularly ejecting particles from its surface~~\cite{45} makes this concept possible.  the frequency, size and velocity of these particle ejections were established to be 1-10 cm in diameter,  6 to 330 cm/s in velocity and were observed to launch episodically with an cumulative frequency of approximately 599 particles over 45 days.  The escape velocity from Bennu is about 20 cm/s, and the time spent by these particles in flight typically is measured in hours before they either escape from gravitational influence, or return to the surface of the asteroid.

\begin{figure}
\centering
\includegraphics[width=0.7\linewidth]{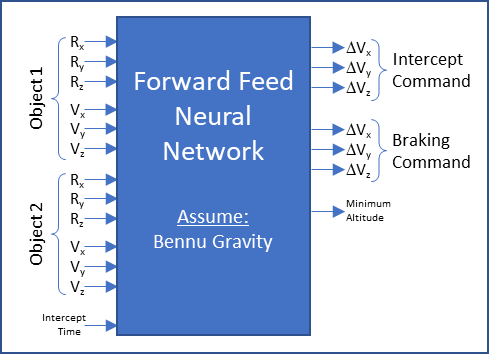}
\caption{Definition of the Problem:  A neural network approach to lambert guidance is defined to reduce processing requirements for a nanospacecraft attempting to rendezvous an ejected particle around Bennu}
\label{guid_fig2}
\end{figure}

In this section, a training set of orbits and their resulting transfer trajectories are established using randomly selected position and velocity parameters around asteroid Bennu as inputs to the Lambert algorithm.  Seven separate filters are trained, one each for three axes of starting velocity, ending velocity plus a minimum transition altitude.  Figure~\ref{guid_fig2}  defines the basic outline of the problem, and shows an example of a transfer trajectory derived using Lambert between two objects in orbit around Bennu.  A Levenburg-Marquardt algorithm is implemented which shows good convergence with a modest number of hidden nodes and layers.  Finally, the resulting trained networks are used to verify performance in an orbital test case.

\subsection{Implementation of Levenburg-Marquardt Optimization}
Research indicates that perhaps the most common training technique for neural networks is the Levenberg-Marquardt algorithm~~\cite{7}.  This method effectively acts as a gradient descent at the beginning of the epoch, and gradually transitions to solve for zero error effectively using a Newton-Raphson style method as the system starts to converge.  Using the Jacobean derived for use in gradient descent, the weight update equation can be reformulated as:

\begin{equation}
     w_i = [P^T P+\mu I]^{-1}J^T(y - y_{train}) + W_{i-1}
\end{equation}

where $\mu$ is a scaler control variable with value $\mu>0$, $w_i$ is the updated weight vector, P is the same Jacobean matrix used in the gradient descent method, $(y-y_{train})$ is the error vector and $w_{i-1}$ the previous weight vector.  This optimizer approximates a small step gradient descent method when $\mu$ is large, but becomes closer to a Newton-Raphson solver as mu approaches zero.  $\mu$ is adjusted during an optimization run, increasing fractionally if the error function increases, but decreasing fractionally when it is converging.  In a nominally converging system, $\mu$ will steadily decrease as the network converges.

\begin{figure}
\includegraphics[width=\linewidth]{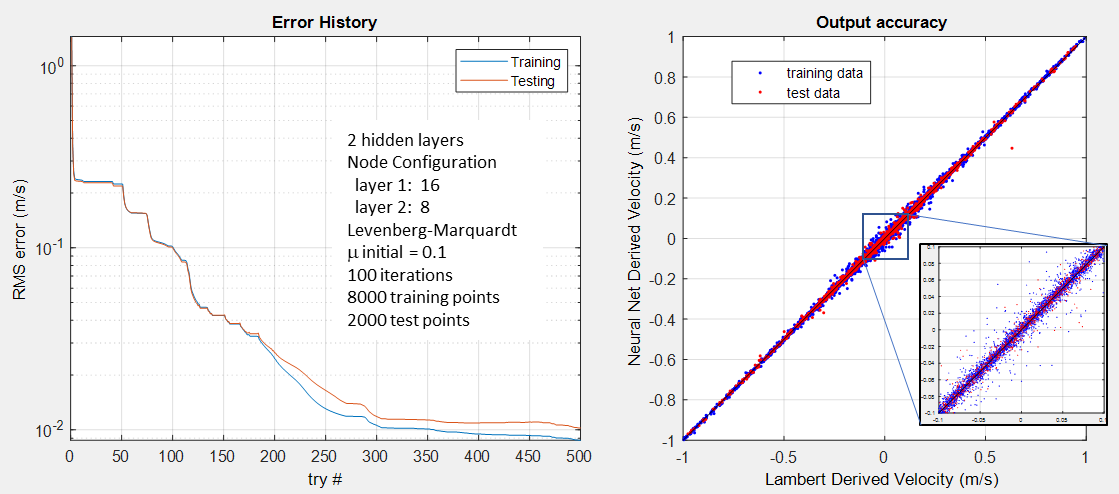}
    \caption{Levenberg-Marquardt Descent for Lambert's solution converges quickly and accurately for the velocity-X axis, and this behavior is near identical for the other two axes, as well as the velocity required at intercept.}
    \label{guid_fig8}
\end{figure}

\begin{figure}
\includegraphics[width=0.7\linewidth]{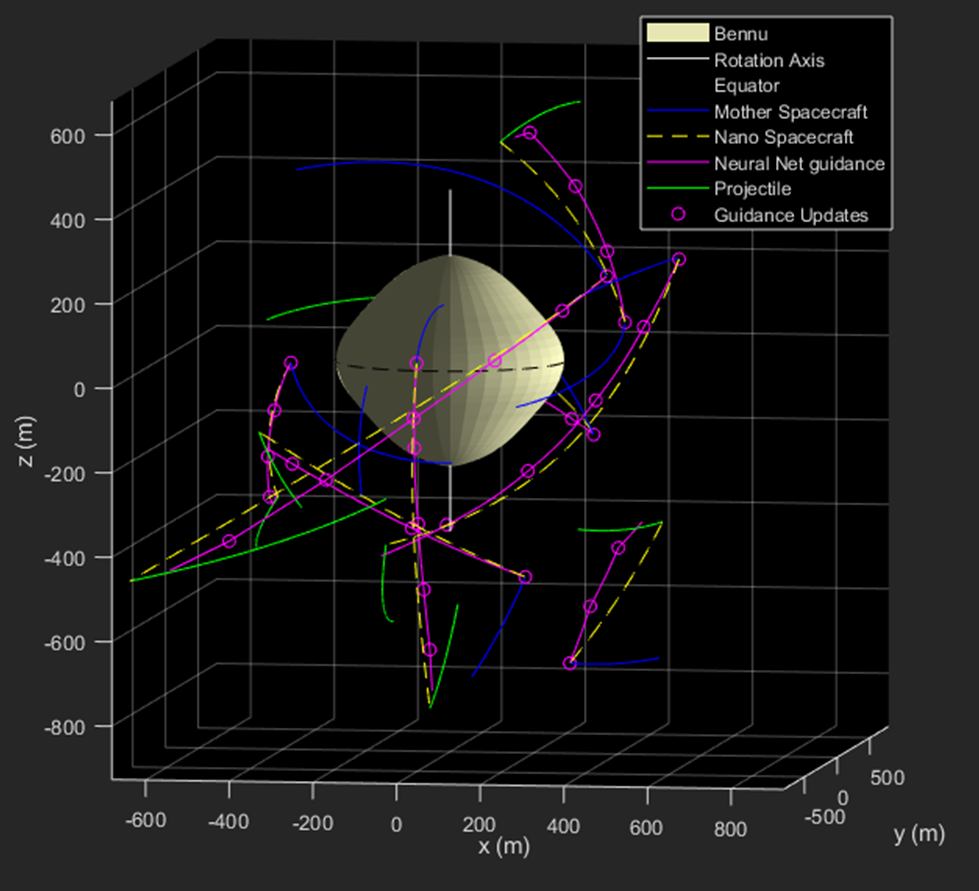}
\centering
    \caption{The overall performance of the resulting neural network guidance is shown, comparing the Lambert derived trajectory (yellow) to the neural network derived trajectory (magenta).  Discrete guidance updates using the neural network are calculated and shown at 1250 second intervals (magenta circles) }
    \label{guid_fig5}
\end{figure}

This algorithm performs well and converges for all six $\Delta V$ estimates.  Figure~\ref{guid_fig8} shows this convergence using this technique for a modest 2 layer system of 16 and 8 hidden nodes respectively.  All other values of $\Delta V$ show similar excellent results, though the estimate for minimum altitude during the intercept trajectory has more difficulty converging.  Performance starts to fall off significantly with less that 16 nodes in the first hidden layer, but is not visibly improved  with increases beyond the nominal values for the two layer system specified as the nominal case.  The rms error of this estimator calculates to 0.9 cm/s, and when implemented as the guidance method, the resulting trajectories can be seen in Figure ~\ref{guid_fig5}

%
%
\subsection{Thruster Configuration Optimizations}
The size and complexity of a spacecraft is significantly effected by its propulsion configuration.  In the case of a nanospacecraft executing the mission of collecting ejected particles from an active asteroid, there would be a requirement for full six degree of freedom motion, and possibly a requirement for redundancy if one of the thrusters fails.  The physical volume required for plumbing, valves and expansion nozzles encourages the designer to reduce the individual number of thrusters to the absolute minimum. This section discusses the minimum configuration needed for full 6 axis control of a nanospacecraft, and then the mathematical techniques to show controllability as well as thruster failure redundancy.  

This section seeks to establish minimal thruster configurations which maximize effectiveness and while still providing single thruster failure redundancy.  The actuation logic for a multiple coupled thruster system can be expressed as:

\begin{equation}
    \begin{bmatrix}
    \boldsymbol{F_b}\\
    \boldsymbol{L_b}\\
   \end{bmatrix} 
    = [\boldsymbol{T}][\boldsymbol{F_t}]
\end{equation}

where $\boldsymbol{F_t}$ is a vector containing the thrust values from each thruster at a given time, $\boldsymbol{F_b}$ is the resulting three dimensional net thrust in the body frame, $\boldsymbol{L_b}$ is the body frame torque vector, and $\boldsymbol{T}$ is the matrix relating thrust values to net forces and torques.

\begin{figure}
\includegraphics[width=\linewidth]{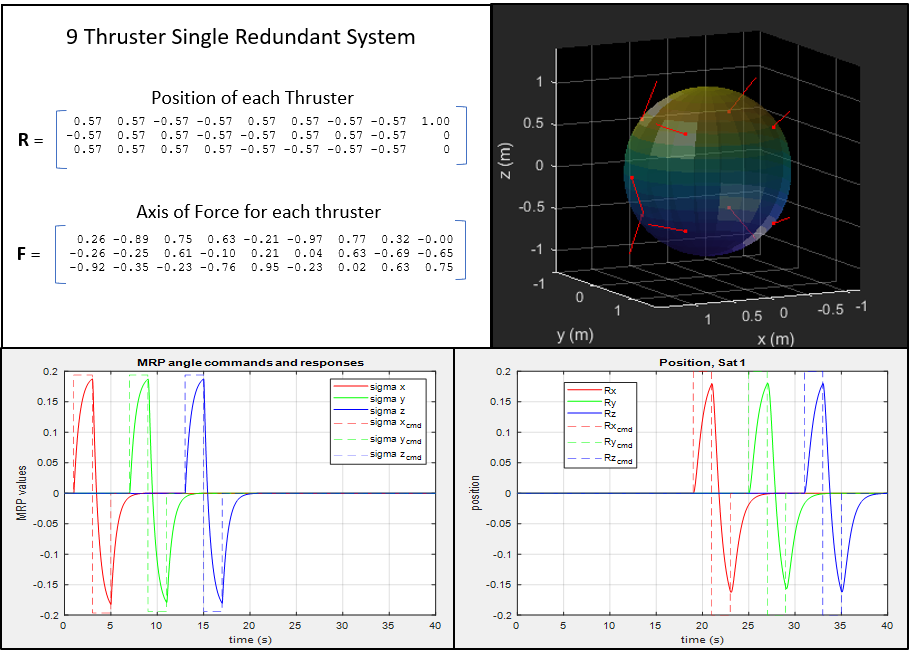}
    \caption{An example of a fully redundant, six degree of freedom thruster configuration derived via pseudo-inverse techniques combined with null vector evaluation.  Improvement of placement and orientation via genetic algorithms provides a path towards optimizing nanospacecraft propulsion systems }
    \label{thrust_fig1}
\end{figure}

In order to get full three dimensional control of a spacecraft in both linear and rotational space, a minimum of seven thrusters is required, with more needed if redundancy is required.  The classical approach to establishing the thruster control levels needed uses the pseudo inverse method such that

\begin{equation}
    \boldsymbol{F_t} =[\boldsymbol{T^{-1}}]
    \begin{bmatrix}
    \boldsymbol{F_b}\\
    \boldsymbol{L_b}\\
   \end{bmatrix} 
\end{equation}

In order this technique to work, there must also be at least one positive definite null space vector to the $\boldsymbol{T}$ matrix, so that thrust levels derived from the pseudo-inverse method above can be adjusted upwards until all are non-zero.  In the case of testing for redundancy, these two conditions must be met for a each subset of the $\boldsymbol{T}$ matrix corresponding to the loss of one thruster.  

Preliminary simulations have established minimal configurations of 9 thrusters which provide full 6 degree of freedom control with single failure redundancy, but these thruster configurations are asymmetric.   Figure~\ref{thrust_fig1} shows an example of such a system configured on a spherical volume which has full six degree of freedom control for both the nominal configuration as well as any given single failures.  Time history performance of this configuration is shown in the bottom half of this same figure.  The proposed dissertation shall execute the analysis necessary to establish full six degree of freedom control with redundancy with a minimum number of thrusters, including placement and orientations optimized for control effectiveness.

\subsection{Generic Capture Mechanism}
\begin{figure}
\includegraphics[width=\linewidth]{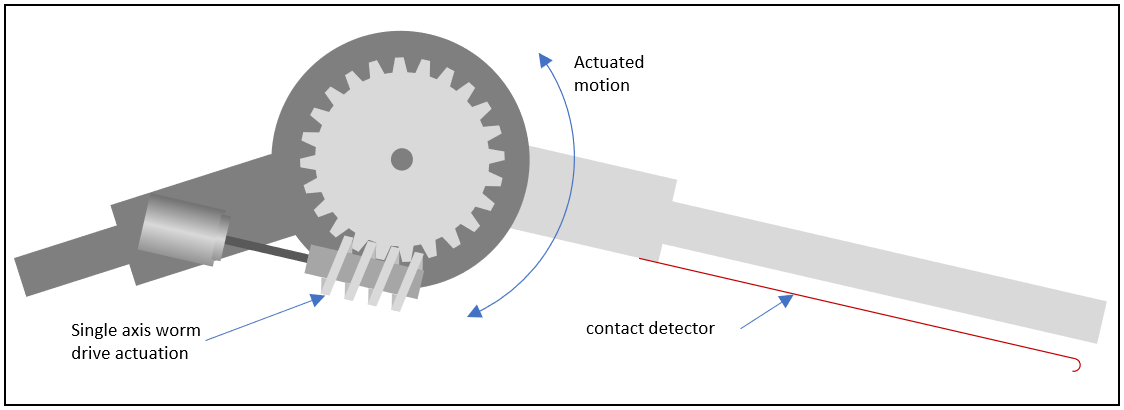}
    \caption{Link Concept:  The candidate link concept uses a miniature worm drive to actuate successive links.  A wire is included in the outline of each link which is triggered upon establishing contact with an object. }
    \label{grappler_fig1}
\end{figure}

This concept involves the deployment of multiple grappling arms, each comprised of identical segments with independently controlled single axis rotational joints.  each link includes   a sensor to establish contact with the object to be grappled.  Figure~\ref{grappler_fig1} provides an overview of one joint of this proposed grappling system.

\begin{figure}
\includegraphics[width=\linewidth]{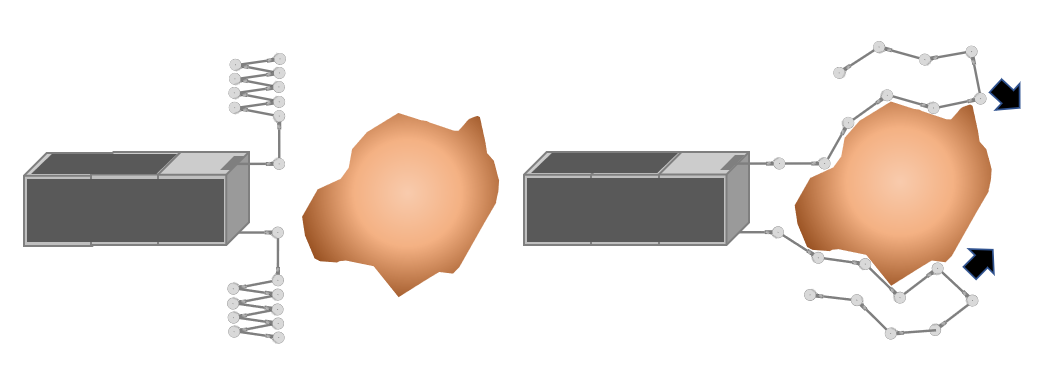}
    \caption{Concept of Operations:  Upon establishing stable proximity with the object to be captured, the spacecraft establishes an approach vector along the spin axis of the object, and deploys the grappler arms one link at a time, where each link stops when sensing contact with the object, and subsequent links then rotate forward, slowly enveloping the object to be captured. }
    \label{grappler_fig2}
\end{figure}

Upon establishing position and rate matching along center of rotation of the object to be grappled, the spacecraft starts to extend the adaptive grappler link by link, stopping each time contact is detected with that link, and then stepping to the next link.  In this fashion, the target object is slowly enveloped in a low-force, controlled fashion until the entire grappling arm is deployed.  This approach is shown in a planar view in Figure~\ref{grappler_fig2}.

The control for this concept needs to be expanded to include the algorithms necessary to establish a rippling motion for the as yet undeployed links so that 1) the amount of energy necessary to deploy the system is reduced, and 2) to reduce the probability that grappler might hit obstructions out beyond the minimal extent of the object being captured.  6-degree of freedom simulation of this concept and the capture sequence will be executed and optimized for system performance leading to a conceptual design.

%
%
\section{Conclusions}
In light of the complexity and risks associated with the sampling experience of  Hayabusa 1 and 2 as well as OSIRIS-REx, establishing alternative approaches to obtain samples from small asteroids is a beneficial exercise.  The discovery of regularly ejected particles from Bennu suggests the possibility that most if not all top shaped near Earth asteroids are active in some similar fashion, enabling the possibility that sampling could be done by capturing these particles in flight and returning them to Earth.  This dissertation proposal documents the development of key technologies enabling this mission. 

Characterisation of the ejected particle environment is established showing that the ejection of surface particles from an active asteroid contributes to the commonly observed 'double-top' shape.  Auto-navigation and Neural network based guidance is then demonstrated.  The concept of achieving auto-navigation using passive line of sight measurements from poorly known objects under the influence of a common gravitational field is quantified, with results showing convergence for both the spacecraft and ejected particle position and velocity estimates.  Significant improvements in performance are plausibly expected when the spacecraft begins maneuver, as observability improves quickly what is effectively temporal triangulation. 

Perceptron based guidance is demonstrated by showing convergence of a trained neural network to mimic the output of the iterative Lambert's solution.  Comparison of trajectories showing both solutions demonstrates the promise of this technique.  Spacecraft control is explored by showing the existence of a 10 thruster placement for redundant 6 degree of freedom control, thus indicating the means to establish optimization. The development of a universal grappling scheme for non-cooperative targets and consistent with known projectile sizes is then explored and control algorithms are developed for its use.  Finally, a demonstration of these technologies is executed on a air-bearing table contoured to mimic the low gravity environment of a small asteroid.  The overall resulting mission design is conceptually designed and risk reduced for key technological components.

\end{document}